\documentclass[fleqn,12pt,twoside]{article}
\usepackage{espcrc1}

\usepackage{graphicx}
\usepackage[figuresright]{rotating}

\def\beq{\begin{equation}}
\def\eeq{\end{equation}}
\def\be{\begin{equation}}
\def\ee{\end{equation}}
\def\beq{\begin{equation}}
\def\eeq{\end{equation}}

\def\as{\alpha_{\mbox{\scriptsize s}}}
\def\eqref#1{(\ref{#1})}
\def\qcold{\hat{q}_{\mbox{\scriptsize{cold}}}}
\def\qhot{\hat{q}_{\mbox{\scriptsize{hot}}}}
\def\EHQ{E_{\mbox{\scriptsize{HQ}}}}

\def\GeV{{\rm Ge\!V}}

\def\fun#1#2{\lower3.6pt\vbox{\baselineskip0pt\lineskip.9pt
  \ialign{$\mathsurround=0pt#1\hfil##\hfil$\crcr#2\crcr\sim\crcr}}}
 \newskip\humongous \humongous=0pt plus 1000pt minus 1000pt

\newcommand{\AmS}{{\protect\the\textfont2
  A\kern-.1667em\lower.5ex\hbox{M}\kern-.125emS}}

\title{Heavy Quarks and QCD Matter}

\author{D. Kharzeev\address[BNL]{Physics Department, \\
        Brookhaven National Laboratory, \\ 
        Upton, New York 11973-5000, USA}%
        \thanks{Invited talk at the International Conference on ``Statistical QCD'', 
Bielefeld, August 26-30, 2001.}}

\begin{document}

\maketitle

\begin{abstract}
I present recent results on the theory of QCD matter production 
in high energy heavy ion collisions and on the interactions 
of heavy quarks in such environment. 
The centrality and rapidity dependence of hadron production is evaluated 
in semi--classical approach. The energy loss of heavy quarks in matter 
is computed.   
The heavy--to--light meson ratio (e.g., $D/\pi$) at moderate transverse 
momenta is demonstrated to be both sensitive to the density of color charges 
in the medium {\it and} infrared stable.    

\end{abstract}

\section{Foreword}

This conference celebrates the contributions made by Helmut Satz to the theory 
of super--dense matter. Helmut was among the very first who started to think 
about the subject at the time of its infancy. His work on QCD matter over 
the years brought the field of statistical QCD to maturity. His 
strong support was vital in establishing experimental heavy ion programs 
at CERN and BNL. All of this has been already 
described at the conference by people 
who have a first--hand knowledge of the history of the field (in which 
the previous Bielefeld 1980 meeting was a major milestone). 
What I would like to add to these accounts is 
Helmut's influence on young physicists: he is, and has always been, an 
inspiration for people entering the field. Many of us, like myself, 
were brought into the field and encouraged by him. Working with Helmut is both enriching 
and enjoyable. I wish new generations of physicists will discover this for themselves. 

\section{Statistical QCD: from small $x$ to high $T$} 

\subsection{Quantum statistics at small $x$}

Small $x$ physics is not usually considered to belong to the realm 
of statistical QCD. Nevertheless, especially at this conference, it is 
worthwhile to emphasize that the concepts of statistical approach 
provide a very useful perspective in small $x$ physics as well. 

Let us begin by noting that a parton fluctuation with a given Bjorken 
$x$ and transverse momentum $k_{\perp}$ inside a hadron with a momentum $P$ 
has, in the Lab frame, a lifetime given 
by the uncertainty relation:
\be
t \sim {x P \over k_{\perp}^2}. \label{life}
\ee
This shows that the partons at larger $x$, and smaller $k_{\perp}$, 
live much longer than partons at small $x$ and large $k_{\perp}$.

Let us denote by $\varphi = \left\{\psi, {\bar{\psi}}, A \right\}$ the set of 
parton fields with $x < x_0$ (with $x_0$ setting some arbitrary 
boundary between ``fast'' and ``slow'' fields), and by $\phi$ the set of the same fields, 
but with $x > x_0$. 
Suppose that we want to compute an expectation 
value of some observable $\cal{O}$.
In doing so, we have to take account of the fact that partons at larger $x > x_0$ 
are effectively ``frozen''; this is done by employing the form familiar 
from the treatment of statistical systems with random, frozen impurities:
\be
\left< \cal{O} \right> = \int D \phi \ \rho(\phi)\  
{{\int D \varphi \ {\cal{O}}(\varphi,\phi) \ \exp(i S(\varphi,\phi)/{\hbar})} \over 
{\int D \varphi \ \exp(i S(\varphi,\phi)/{\hbar})}}, \label{genf} 
\ee 
where $S$ is the action, $\rho(\phi)$ describes the distribution of large $x$ partons, 
and we have explicitly written down the Planck 
constant $\hbar$.
The meaning of (\ref{genf}) is simple -- the ``frozen'' fields $\phi$ are not 
a dynamical part of the system of the ``fast'' fields $\varphi$; rather, they act as 
impurities, or sources. This formulation is the basis for McLerran--Venugopalan 
model of hadron structure at small $x$; since glasses are among the physical 
systems with large relaxation time, one may also call such system 
a ``color glass condensate'' (see \cite{LM,RV} and references therein). 
Renormalization group equations with respect to the changing scale $x_0$ allow then 
to reconstruct QCD evolution: partons radiated by sources at larger $x$ themselves 
become sources for radiation at even smaller $x$.

At sufficiently small $x$ and/or large atomic number of the nucleus, 
the density of partons will become very large and 
the system will thus cease to be dilute. 
What will it look like? What kind of dynamics will govern its properties? 
To address these questions, let us first note that for a system with 
large number of gluons the action is large, $S \gg \hbar$. Such systems 
are appropriately described by using the semi--classical approximation. 
To go further, we need to establish the dependence of the action on the 
coupling constant. To do this, let us re-scale the gluon fields 
in the QCD Lagrangian as follows: $A_{\mu}^a \to \tilde{A}_{\mu}^a = 
g A_{\mu}^a$. In terms of new fields, $\tilde{G}_{\mu \nu}^a = 
g G_{\mu \nu}^a = \partial_{\mu} \tilde{A}_{\nu}^a - \partial_{\nu} 
\tilde{A}_{\mu}^a +  f^{abc} \tilde{A}_{\mu}^b \tilde{A}_{\nu}^c$, 
and the dependence of the action on the coupling constant is given by  
\beq
S \sim \int {1 \over g^2}\ \tilde{G}_{\mu \nu}^a  \tilde{G}_{\mu \nu}^a 
\ d^4 x. \label{act}
\eeq
Let us now consider a classical configuration of gluon fields; by definition, 
$\tilde{G}_{\mu \nu}^a$ in such a configuration does not depend on 
the coupling, and the action is large, $S \gg \hbar$. The number of 
quanta in such a configuration is then
\beq
N_g \sim {S \over \hbar} \sim {1 \over \alpha_s}\ \rho_4 V_4, \label{numb}
\eeq
where we re-wrote (\ref{act}) as a product of four--dimensional 
action density $\rho_4$ and the four--dimensional volume $V_4$. 
 
The effects of non--linear interactions among the gluons become 
important when $\partial_{\mu} \tilde{A}_{\mu} \sim \tilde{A}_{\mu}^2$ 
(this condition can be made explicitly gauge invariant if we derive it 
from the expansion of a correlation function of gauge-invariant 
gluon operators, e.g., $\tilde{G}^2$). In momentum space, this 
equality corresponds to 
\beq
Q_s^2 \sim \tilde{A}^2 \sim (\tilde{G}^2)^{1/2} = 
\sqrt{\rho_4}; \label{nonlin}
\eeq
$Q_s$ is the typical value of the gluon momentum below which 
the interactions become essentially non--linear. 

Consider now a nucleus $A$ boosted to a high momentum. By uncertainty 
principle, the gluons with transverse momentum $Q_s$ are extended 
in the longitudinal and proper time directions by $\sim 1/Q_s$; 
since the transverse area is $\pi R_A^2$, the four--volume 
is $V_4 \sim \pi R_A^2 / Q_s^2$. The resulting four--density from 
(\ref{numb}) is then 
\beq
\rho_4 \sim \alpha_s\ {N_g \over V_4} \sim \alpha_s\ {N_g\ Q_s^2 
\over \pi R_A^2} 
\sim Q_s^4, \label{class}
\eeq
where at the last stage we have used the non--linearity condition (\ref{nonlin}),  
$\rho_4 \sim Q_s^4$. 

Identifying the number of gluons in the 
infinite momentum frame with the gluon structure function $N_g \sim x G(x,Q_s^2)$, we arrive at the 
condition 
\beq 
Q_s^2 \sim  \alpha_s \ {x G_A(x,Q_s^2) \over \pi R_A^2}, \label{qsat}
\eeq
originally derived in \cite{GLR,MUQI,BM} as 
the criterion for ``parton saturation'' (for a discussion of saturation 
in terms of the partons in the final state, see \cite{KJE,VR}). 
This simple derivation \cite{DK1}  
illustrates that the physics in the high--density regime can potentially 
be understood in terms of classical gluon fields. 

\subsection{Classical QCD and particle production in heavy ion collisions}

The energy dependence of saturation scale 
$Q_s$ is determined by the 
$x-$ dependence of the gluon structure function (see (\ref{qsat})).  
 In spite of significant uncertainties in determination 
of the gluon structure functions, the following observation 
\cite{GW} is very important: the  
HERA data exhibit scaling when plotted as a function of variable 
\beq
\tau \,=\, {Q^2 \over Q_0^2} \ \left({x \over x_0}\right)^{\lambda}, 
\eeq
where $\lambda \simeq 0.25 \div 0.3$. 
In saturation scenario, this scaling translates in the following $x$ dependence 
of dimensionful scale $Q_s$: 
\beq
Q_s^2(x) = Q_0^2 \ (x_0 / x)^{\lambda}. \label{xdep}
\eeq
Since the rapidity $y$ and Bjorken variable are related by $\ln 1/x = y$, 
(\ref{xdep}) leads to the  
dependence of the saturation scale $Q_s^2$ on rapidity:
\beq
Q_s^2(s; \pm y) = Q_s^2(s; y = 0)\ \exp(\pm \lambda y). \label{qsy}
\eeq

\begin{figure}[htb]
\begin{minipage}[t]{80mm}
\includegraphics[width=14pc]{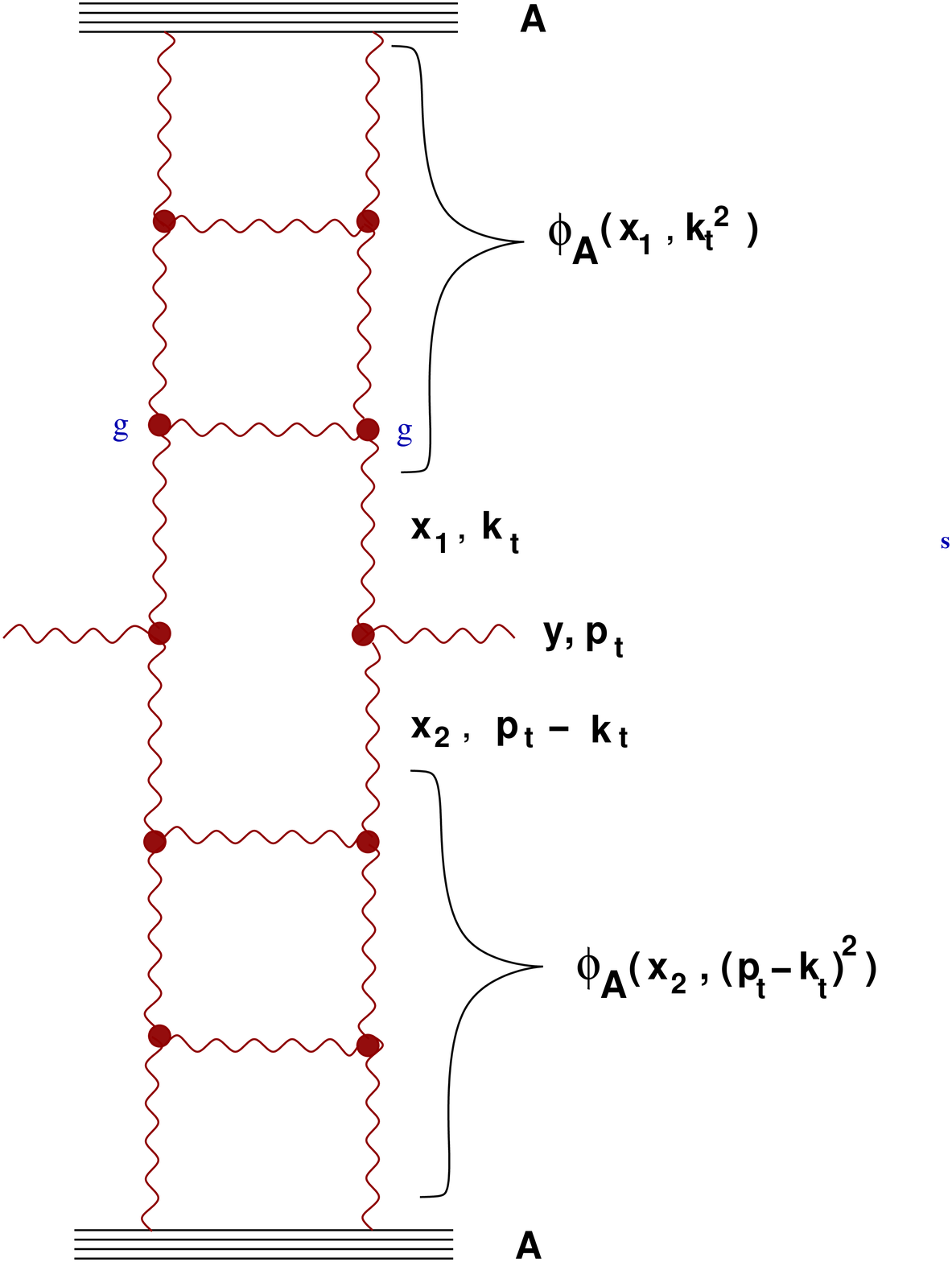}
\caption{ Mueller diagram of classical gluon emission.}
\label{emis}
\end{minipage}
\end{figure}

Let us now evaluate the rapidity and centrality dependences of hadron production in 
heavy ion collisions basing on this picture 
\cite{KN,KL}. 
We need to evaluate the leading tree diagram describing 
emission of gluons on the classical level, see Fig. \ref{emis}.
 
To do this, we introduce the unintegrated gluon distribution $\varphi_A (x, k_t^2)$ which 
describes the probability to find a gluon with a given $x$ and transverse 
momentum $k_t$ inside the nucleus $A$. As follows from this definition, 
the unintegrated distribution is related to the gluon structure function by
\beq
xG_A(x, p_t^2) = \int^{p_t^2} d k_t^2 \ \varphi_A(x, k_t^2);
\eeq
when $p_t^2 > Q_s^2$, the unintegrated distribution corresponding to the bremsstrahlung 
radiation spectrum is 
\beq
\varphi_A(x, k_t^2) \sim {\alpha_s \over \pi} \ {1 \over k_t^2}.
\eeq 
In the saturation region, the unintegrated gluon distribution has only logarithmic dependence on the 
transverse momentum: 
\beq
\varphi_A(x, k_t^2) \sim {S_A \over \alpha_s}; \ k_t^2 \leq Q_s^2, \label{unint}
\eeq
where $S_A$ is the nuclear overlap area, determined by the atomic numbers of the 
colliding nuclei and by centrality of the collision.

 The differential cross section 
of gluon production in a $AA$ collision can now be written down as \cite{GLR,GM}
\beq
E {d \sigma \over d^3 p} = {4 \pi N_c \over N_c^2 - 1}\ {1 \over p_t^2}\ \int d k_t^2 \ 
\alpha_s \ \varphi_A(x_1, k_t^2)\ \varphi_A(x_2, (p-k)_t^2), \label{gencross}    
\eeq
where $x_{1,2} = (p_t/\sqrt{s}) \exp(\pm \eta)$, with $\eta$ the (pseudo)rapidity of the 
produced gluon; the running coupling $\alpha_s$ has to be evaluated at the 
scale $Q^2 = max\{k_t^2, (p-k)_t^2\}$. 
The rapidity density is then evaluated from (\ref{gencross}) according to 
\beq
{dN \over d y} = {1 \over \sigma_{AA}}\ \int d^2 p_t \left(E {d \sigma \over d^3 p}\right), 
\label{rapden}
\eeq
where $\sigma_{AA}$ is the inelastic cross section of nucleus--nucleus interaction. 
\begin{figure}[htb]
\begin{minipage}[t]{80mm}
\includegraphics[width=18pc]{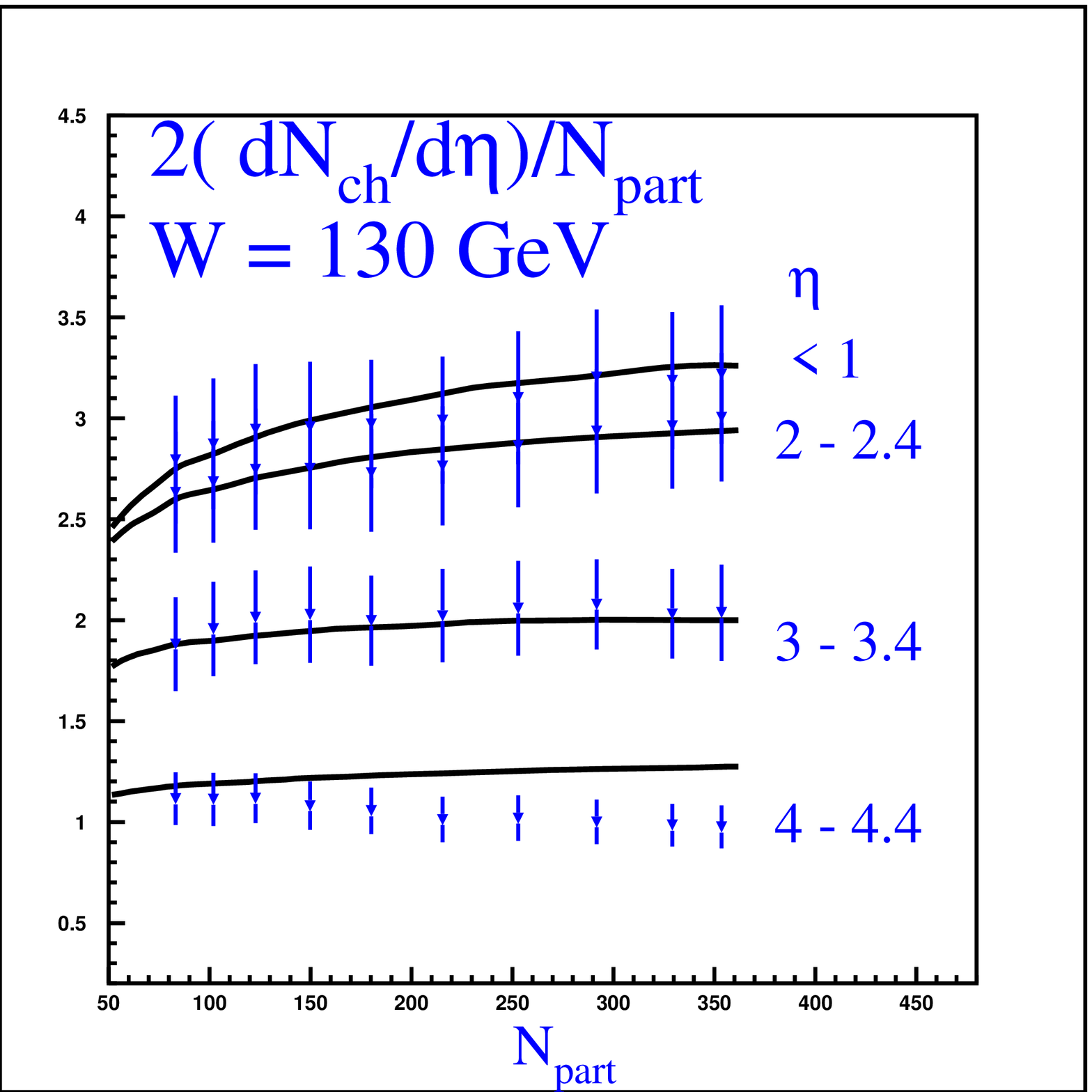}
\caption{ Centrality dependence of charged hadron production per participant at different 
pseudorapidity $\eta$ intervals in $Au-Au$ collisions 
at $\sqrt{s} = 130$ GeV, from \cite{KL}; the data are from \cite{PHOBOS130}.}
\label{fig1}
\end{minipage}
\hspace{\fill}
\begin{minipage}[t]{75mm}
\includegraphics[width=18pc]{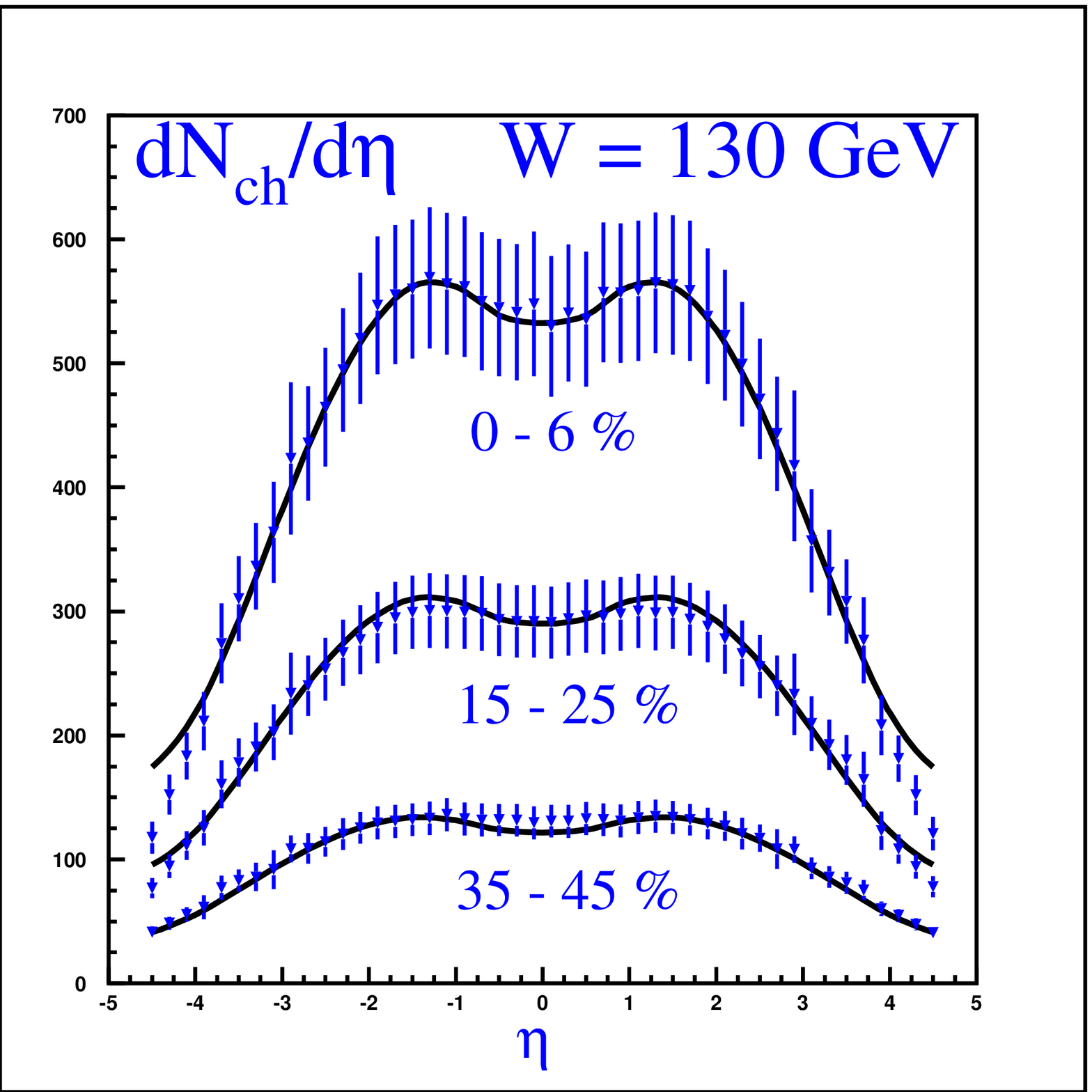}
\caption{ Pseudo--rapidity dependence of charged hadron production at different cuts on centrality 
in $Au-Au$ collisions 
at $\sqrt{s} = 130$ GeV, from \cite{KL}; the data are from \cite{PHOBOS130}.}
\label{fig2}
\end{minipage}
\end{figure}
Evaluation of Eqs.(\ref{gencross}) and (\ref{rapden}) leads to the 
following simple analytical formula \cite{KL}, which exhibits the scaling properties of 
hadron multiplicity in nucleus--nucleus collisions:

\beq
{dN \over d y} = c\ N_{part}\ \left({s \over s_0}\right)^{\lambda \over 2}\ e^{- \lambda |y|}\ 
\left[\ln\left({Q_s^2 \over \Lambda_{QCD}^2}\right) - \lambda |y|\right]\ 
\left[ 1 +  \lambda |y| \left( 1 - {Q_s \over \sqrt{s}}\ e^{(1 + \lambda/2) |y|} \right)^4 \right],  
\label{finres}
\eeq
with $Q_s^2(s) = Q_s^2(s_0)\ (s /s_0)^{\lambda / 2}$.
This formula expresses the predictions of 
high density QCD for the energy, centrality, rapidity, and atomic number dependences 
of hadron multiplicities in nuclear collisions in terms of a single scaling function. 
Once the energy--independent constant $c \sim 1$ and $Q_s^2(s_0)$ are determined 
at some energy $s_0$, Eq. (\ref{finres}) contains no free parameters. 

The results for the $Au-Au$ collisions at $\sqrt{s} = 130$ GeV based on Eq.(\ref{finres}) 
are presented in Figs \ref{fig1} 
and \ref{fig2}. One can see that the agreement with the data is quite good. If it persists 
at higher energies, one may conclude that parton saturation indeed adequately describes 
the initial conditions created in relativistic heavy ion collisions. 
Since saturation provides good conditions for parton thermalization \cite{AM}, 
we may expect that the final goal of producing the equilibrated QCD matter in the laboratory 
may be within reach. We thus have to look for the probes which can be 
used for its diagnostics. One, by now famous, 
probe of QCD matter is the heavy quarkonium \cite{MS}. Another probe is provided 
by high $p_t$ jets (see \cite{XNW}).
 I am now going to discuss a recent proposal, involving heavy quarks at high $p_t$ \cite{DK}.

\section{Heavy quark energy loss in QCD matter}

Let us begin by recalling the basic features of gluon radiation caused by
propagation of a fast parton (quark) through QCD medium.
As was pointed out in \cite{BDPS}, the accompanying radiation is
determined by multiple rescattering of the radiated gluon in the
medium.  The gluon, during its formation time given again by (\ref{life}) 
\beq 
   t_{form} \simeq \frac{\omega}{k_{\perp}^2}\,,
\label{form}
\eeq
accumulates a typical transverse momentum
\beq 
   k_{\perp}^2 \simeq \mu^2 \ {t_{form} \over \lambda}, \label{walk}
\eeq
with $\lambda$ the 
mean free path and $\mu^2$ the characteristic momentum transfer
squared in a single scattering.  This is the random walk pattern with
an average number of scatterings given by the ratio
$t_{form}/\lambda$.

Combining (\ref{walk}) and (\ref{form}) we obtain
\beq\label{Ncoh}
  N_{coh}=\frac{t_{form}}{\lambda} = \sqrt{\frac{\omega}{\mu^2\,\lambda}}
\eeq
describing the number of scattering centers which participate, {\em
coherently}, in the emission of the gluon with a given energy
$\omega$.  For sufficiently large gluon energies,
$\omega>\mu^2\lambda$, when the coherent length exceeds the mean free
path, $N_{coh}>1$.  In this situation the standard Bethe-Heitler
energy spectrum per unit length describing {\em independent}\/
emission of gluons at each center gets suppressed:
\beq\label{spec}
 \frac{dW}{d\omega dz} = \left(
\frac{dW}{d\omega dz}\right)^{\mbox{\scriptsize BH}} \cdot\frac1{N_{coh}}
= \frac{\as C_R}{\pi\omega\,\lambda}\cdot\sqrt{\frac{\mu^2\,\lambda}{\omega}} 
=  \frac{\as C_R}{\pi\omega} \sqrt{\frac{\hat{q}}{\omega}}.
\eeq
Here $C_R$ is the ``color charge'' of the parton projectile
($C_R=C_F=\frac{N_c^2 - 1}{2 N_c} = 4/3$ for the quark case we are
interested in).

In \eqref{spec} we have substituted the characteristic ratio 
$\mu^2/\lambda$ by the so-called gluon 
{\em transport coefficient}\/
\cite{BDMPS1}
\beq
 \hat{q} \equiv 
 \rho \ \int {d \sigma \over dq^2}\ q^2\ dq^2, \label{qhat}
\eeq
which is proportional to the density $\rho$ of the scattering centers
in the medium and describes the typical momentum transfer in the gluon
scattering off these centers.

The transport coefficient 
for cold nuclear matter was expressed in \cite{BDMPS1} as
\beq 
\hat{q} \simeq {4 \pi^2 \alpha_s N_c\over N_c^2-1}\ 
\rho\ [xG(x, Q^2)], \label{qhatcold}
\eeq
with $\rho \simeq 0.16\ {\rm fm^{-3}}$  the average nuclear
density and $[xG(x)]$ the gluon density in a nucleon. 
Taking $\alpha_s \simeq 0.5$ and $[xG(x)] \simeq 1$ (at
$x < 0.1$),  yields 
\beq
 \qcold \simeq 0.01 \ {\rm GeV^3} \simeq 8 \ \rho. \label{numqhat}
\eeq  
This estimate  
is an agreement with 
the result of the analysis of the gluon $p_{\perp}$ broadening from
the experimental data on $J/\psi$ transverse momentum distributions
\cite{KNS}, which in the present notation yielded
\beq 
\hat{q} = (9.4 \pm 0.7) \> \rho\,.
\eeq
An estimate \cite{BDMPS1} for a hot medium based on perturbative
treatment of gluon scattering in quark--gluon plasma with $T\sim
250$~MeV resulted in the value of the gluon transport coefficient of
about factor {\em twenty}\/ larger than \eqref{numqhat}:
%
\beq
   \qhot \simeq 0.2\ \rm{GeV}^3\>\simeq\> 20\, \qcold \,. 
\label{qhathot}
\eeq

Multiplying \eqref{spec} by the length $L$ of the medium
traversed,\footnote{For the sake of simplicity we assume here that the
medium is static and uniform.}  we arrive at the following expression
for the inclusive energy distribution of gluons radiated by a quark:
\beq\label{omega1}
  \frac{dW}{d\omega } \simeq \frac{\as C_F}{\pi\,\omega}
  \sqrt{\frac{\omega_1}{\omega}}, \qquad \omega \><\> 
  \omega_1\equiv  \hat{q}L^2\,.
\eeq 

The fact that the medium induced radiation vanishes for
$\omega>\omega_1$ has a simple physical explanation, as according to
\eqref{Ncoh} the formation time of such gluons starts to exceed the
length of the medium:
\[
 t_{form}= \lambda\cdot\sqrt{\frac{\omega}{\mu^2\lambda}} 
= \sqrt{\frac{\omega}{\hat{q}}} = L\cdot
\sqrt{\frac{\omega}{\omega_1}} \>>\> L\,.
\] 

Another important feature of medium induced radiation is the relation
between the transverse momentum and the energy of the emitted gluon.
Indeed, from \eqref{form} and \eqref{walk} (see also \eqref{qhat}) we
derive
\beq
  k_{\perp}^2 \simeq \sqrt{\hat{q}\ \omega} \label{star}.
\eeq 
This means that the angular distribution of gluons with a given energy
$\omega$ is concentrated at a characteristic energy- (and medium-)
dependent emission angle
\beq\label{angle}
\theta \simeq \frac{k_\perp}{\omega} 
\sim \left(\frac{\hat{q}}{\omega^3}\right)^{1/4}. 
\eeq


Gluon bremsstrahlung off a heavy quark differs from the case of a
massless parton (produced in a process with the same hardness scale)
in one respect: gluon radiation is suppressed at angles smaller than
the ratio of the quark mass $M$ to its energy $E$.  
Indeed, the distribution of soft gluons radiated by a heavy quark is
given by
\beq
dP  = {\as\ C_F \over \pi}\ 
{d\omega \over \omega}\ {k_{\perp}^2 \,dk_\perp^2\over 
(k_{\perp}^2 + \omega^2 \theta_0^2)^2}, \qquad
\theta_0\equiv\frac{M}{E}\,,
\label{dist}
\eeq
where 
the strong coupling constant $\alpha_s$ should be evaluated at the scale
determined by the denominator of (\ref{dist}).  Equating, in the
small-angle approximation, $k_\perp$ with $\omega\theta$ we
conclude that the formula (\ref{dist}) differs from the standard
bremsstrahlung spectrum
\beq
 dP_0\> \simeq \> \frac{\as\,C_F}{\pi}
 \frac{d\omega}{\omega}\,\frac{dk_\perp^2}{k_\perp^2}
\> =\> \frac{\as\,C_F}{\pi} \frac{d\omega}{\omega}\,\frac{d\theta^2}{\theta^2}
\eeq
by the factor 
\beq\label{factor}
 dP_{\mbox{\scriptsize HQ}} = dP_0\cdot \left( 
1+\frac{\theta_0^2}{\theta^2}\right)^{-2}
\eeq

This effect is known as the ``dead cone'' phenomenon.  Suppression of
small-angle radiation has a number of interesting implications, such
as perturbative calculability of (and non-perturbative $\Lambda/M$
corrections to) heavy quark fragmentation functions~\cite{DKT,NW},
multiplicity and energy spectra of light particles accompanying hard
production of a heavy quark~\cite{hqmulspec}.

In the present context we should compare the angular distribution
of gluons induced by the quark propagation in the medium with the
size of the dead cone.
To this end, for the sake of a semi-quantitative estimate, we 
substitute the characteristic angle \eqref{angle} into 
the dead cone suppression factor \eqref{factor} and combine it with
the radiation spectrum \eqref{spec} to arrive at
\beq
I(\omega) = \omega {d W \over d \omega} = {\alpha_s\ C_F \over \pi}
\sqrt{{\omega_1 \over 
\omega}} \ {1 \over (1 + (\ell\, \omega)^{3/2})^2}, 
\label{eq:spechq}
\eeq 
where
\beq
  \ell \equiv  \hat{q}^{-1/3}\ \left({M \over E}\right)^{4/3}. \label{apar}
\eeq

To see whether the finite quark mass essentially affects the medium
induced gluon yield, we need to estimate the product $\ell\omega$ for
the maximal gluon energy $\omega\simeq\omega_1$ to which the original
distribution
\eqref{spec} extends:
\beq
 \ell\omega_1= \hat{q}^{-1/3}\ \left({M \over E}\right)^{4/3}\cdot
\hat{q}L^2
= \left(\frac{\EHQ}{E}\right)^{4/3}\,, \qquad \EHQ\equiv M\sqrt{\hat{q}L^3}.
\eeq
This shows that the quark mass becomes irrelevant when the quark
energy exceeds the characteristic value $\EHQ$ which depends on the
size of the medium and on its ``scattering power'' embodied into the
value of the transport coefficient.

Which regime is realized in the experiments on heavy quark 
production in nuclear collisions? 
Taking $M = 1.5\ \rm{GeV}$ for charm quarks and using the values
\eqref{numqhat} and \eqref{qhathot} 
we estimate
\begin{eqnarray}
\label{EHQc}
\EHQ &=& \sqrt{\qcold}\ L^{3/2}\ M \simeq \ 20\ {\GeV}
\left({L \over 5\, {\rm fm}}\right)^{3/2}, \\
\label{EHQh}
\EHQ &=& \sqrt{\qhot}\ L^{3/2}\ M \simeq \ 
92\  {\GeV} 
\left({L \over 5\, {\rm fm}}\right)^{3/2},
\end{eqnarray}
for the cold and hot matter, respectively.  We observe that for the
transverse momentum (energy) distributions of heavy mesons the
inequality $E \ll \EHQ$ always holds in practice, especially for the
hot medium.  We thus conclude that the pattern of medium induced gluon
radiation appears to be {\em qualitatively different for heavy and
light quarks}\/ in the kinematical region of practical interest.


The issue of in-medium quenching of inclusive particle spectra was
recently addressed in \cite{BDMSquen}.  The $p_{\perp}$ spectrum is
given by the convolution of the transverse momentum distribution in an
elementary hadron--hadron collision, 
evaluated at a shifted value $p_\perp+\epsilon$, with the 
distribution $D(\epsilon)$ in the energy $\epsilon$ lost 
by the quark to the medium-induced gluon radiation:
\beq
{d \sigma^{med} \over d p_{\perp}^2} = \int d \epsilon \ D(\epsilon)\ 
{d \sigma^{vac} \over d p_{\perp}^2}( p_{\perp} + \epsilon )
\equiv {d \sigma^{vac} \over d p_{\perp}^2}( p_{\perp}) \cdot Q(p_\perp),
  \label{defpt}
\eeq
with $Q(p_\perp)$ the medium dependent {\em quenching factor}.  The
two facts, namely that in the essential region $\epsilon\ll p_\perp$
and that the vacuum cross section is a steeply falling function, allow
one to simplify the calculation of the quenching factor $Q$ by
adopting the exponential approximation for the $\epsilon$-integral in
\eqref{defpt}:
\beq\label{Qfint}
 Q(p_\perp) \simeq \int d \epsilon \ D(\epsilon)\ \exp\left\{ 
\frac{\epsilon}{p_\perp} \cdot {\cal{L}} \right\} \,, \qquad 
{\cal{L}} \equiv \frac{d}{d\ln p_\perp} 
\ln \left[ {d \sigma^{vac} \over d p_{\perp}^2}( p_{\perp})\right].
\eeq 
This integral results in the {\em Mellin moment}\/ of the quark
distribution,  
\beq\label{Qexp}
Q(p_\perp)\>=\> 
\tilde{D}(\nu) = \exp\left[ - \nu\ \int_0^{\infty} d \omega \ 
 N(\omega)\ e^{-\nu \omega}\right], \quad \nu=\frac{{\cal{L}}}{p_\perp},
\label{tildf}
\eeq  
where $N(\omega)$ is the {\em integrated gluon multiplicity}\/ defined
according to (see \cite{BDMSquen} for details) 
\beq
N(\omega) \equiv \int_{\omega}^{\infty} d\omega' \, {d W (\omega') \over d
\omega'}\,.
 \label{intm}
\eeq

The use of (\ref{Qexp}) furnishes our final result:
\beq
Q_H(p_{\perp}) \simeq \exp \left[- {2 \alpha_s C_F \over \sqrt{\pi}}\ 
L\,\sqrt{\hat{q}\frac{{\cal{L}}_H}{p_\perp}}
 + 
{16 \alpha_s C_F \over 9 \sqrt{3}} L
\left( \frac{ \hat{q}\> \> M^2}{M^2+p_\perp^2}\right)^{1/3}  \right].
\label{finres1}
\eeq
The first term in the exponent in (\ref{finres1}) represents the
quenching of the transverse momentum spectrum which is universal for
the light and heavy quarks,
(modulo the difference of the ${\cal{L}}$ parameters 
determined by the $p_{\perp}$ distributions in the vacuum).  The second
term is specific for heavy quarks. It has a positive sign, which means
that the suppression of the heavy hadron $p_{\perp}$ distributions is
always smaller than that for the light hadrons. This is a
straightforward consequence of the fact that the heavy quark mass
suppresses gluon radiation. At very high transverse momenta, both
terms vanish -- this is in accord with the QCD factorization theorem,
stating that the effects of the medium should disappear as $p_{\perp}
\to \infty$. How fast this regime is approached depends, however, on
the properties of the medium encoded in the value of the transport
coefficient $\hat{q}$ and in the medium size $L$.

Constructing the ratio of the quenching functions, we estimate the
heavy-to-light enhancement factor as
\begin{equation}
\frac{Q_H(p_{\perp})}{Q_L(p_{\perp})} \>\simeq\> 
 \exp \left[ {16 \alpha_s C_F \over 9 \sqrt{3}} L
\left( \frac{ \hat{q}\> \> M^2}{M^2+p_\perp^2}\right)^{1/3}  \right].
\label{eq:ratio}
\end{equation}
Basing on (\ref{eq:ratio}) we find \cite{DK} that hot QCD matter leads 
to a strong, factor of $2\div3$, medium--dependent enhancement of the heavy quark yields 
with respect to the yield of light quarks at moderately large $p_t > M$. 
Experimentally, this effect should 
manifest itself as an enhancement of the heavy--to--light ratios such as 
$D/\pi$. It will also be of interest to study the $B/D$ ratio.

\vskip0.3cm
I wish to thank Helmut Satz for inspiring my interest in the topics discussed in this talk.   
I am grateful to my collaborators Yuri Dokshitzer, Eugene Levin and Marzia Nardi, 
with whom the results presented above were obtained. I am indebted to
 Rolf Baier, Jean-Paul Blaizot, Kari Eskola, Miklos Gyulassy, 
Keijo Kajantie, Larry McLerran, Al Mueller, Berndt M{\"u}ller, Vesa Ruuskanen, 
Xin-Nian Wang, and Raju Venugopalan for very useful discussions. 
This work was supported by the U.S. Department of Energy under Contract No. 
DE-AC02-98CH10886.

\end{document}